\documentclass[12pt]{iopart}
\begin{document}

\comment[On a CFT prediction]{On a CFT prediction in the sine-Gordon model}

\author{Carlos Na\'on\dag\ \footnote[3]{To
whom correspondence should be addressed (naon@fisica.unlp.edu.ar)}  and Mariano
Salvay\dag}

\address{\dag\ Instituto de F\'{\i}sica La Plata, Departamento de F\'{\i}sica, Facultad de Ciencias Exactas,
Universidad Nacional de La Plata, C.C. 67, La Plata 1900, Argentina}

\begin{abstract}
A quantitative prediction of Conformal Field Theory (CFT), which relates the second
moment of the energy-density correlator away from criticality to the value of the
central charge, is verified in the sine-Gordon model. By exploiting the boson-fermion
duality of two-dimensional field theories, this result also allows to show the
validity of the prediction in the strong coupling regime of the Thirring model.

\end{abstract}

%Uncomment for PACS numbers title message

% Uncomment for Submitted to journal title message
%\submitto{\JPA}

% Comment out if separate title page not required
\maketitle

\nosections Some time ago Cardy \cite{Cardy} derived a quantitative prediction of
conformal invariance \cite{BN,DMS} for 2D systems in the scaling regime, away from the
critical point. Starting from the so called 'c-theorem' \cite{Zamolodchikov}, he was
able to relate the value of the conformal anomaly $c$, which characterizes the model
at the critical point, to the second moment of the energy-density correlator in the
non-critical theory:
\begin{equation} \label{1}
\int d^2x\, |x|^2\, \langle\varepsilon(x)\varepsilon(0)\rangle = \frac{c}{3\, \pi
\,t^2\,(2 - \Delta_\varepsilon)^2} ,\end{equation} where $\varepsilon$ is the
energy-density operator, $\Delta_\varepsilon$ is its scaling dimension and
$t\propto(T-T_c)$ is the coupling constant of the interaction term that takes the
system away from criticality. It is interesting to notice that a similar sum rule has
been recently obtained by Jancovici \cite{Jancovici} in the context of Classical
Statistical Mechanics. This author considered the correlations of the number-density
of particles in a 2D two-component plasma (Coulomb gas).

The validity of (\ref{1}) has been explicitly verified for the Ising model
\cite{Cardy}, for 2D self-avoiding rings \cite{Cardy2}, and for the Baxter model
\cite{Naon}. In this last case the formula could be checked only in the weak-coupling
limit, by describing the system in terms of a massive Thirring model \cite{LP} and
performing a first-order perturbative computation.

The main purpose of this comment is to show that (\ref{1}) also holds for a bosonic
QFT with highly non-trivial interactions, the well-known sine-Gordon (SG) model with
Euclidean Lagrangian density given by
\begin{equation} \label{2} L = \frac{1}{2}(\partial_{\mu}\Phi)^2 -
\frac{\alpha}{\lambda}\cos(\sqrt{\lambda}\Phi) + \frac{\alpha}{\lambda},
\end{equation} where $\alpha$ and $\lambda$ are real constants.

In this work we shall perform a perturbative computation up to second order in
$\lambda$. We are then naturally led to consider the renormalization of this theory.
Fortunately this issue has been already analyzed by many authors
\cite{Coleman,Samuel,Schroer,Banks,Wiegmann,Amit}. One of the main conclusions is that
for $\lambda < 8 \pi$ a normal order procedure that eliminates the contributions of
the tadpoles is enough to have a finite theory. The only effect of this prescription
is to renormalize the constant $\alpha$. We shall then restrict our study to this
case.

Since we want to verify equation (\ref{1}) for the model given by (\ref{2}), we will
take $\varepsilon=(1-\cos\sqrt{\lambda}\Phi)/\lambda$ and $t=\alpha$. Taking into
account that the model of free massless scalars has a conformal charge $c=1$ and that
the scaling dimension of $\varepsilon$ for this case is equal to $\lambda/4\pi$,
(\ref{1}) reads
\begin{equation} \label{3}
F(\alpha,\lambda)=\int d^2x\, |x|^2\, \langle\varepsilon(x)\varepsilon(0)\rangle =
\frac{1}{3\, \pi \,\alpha^2\,(2 - \frac{\lambda}{4\pi})^2}. \end{equation}

Expanding the interaction term up to order $\lambda^2$ one has $\varepsilon=
\frac{1}{2}\Phi^2 - \frac{\lambda}{4!}\Phi^4 + \frac{\lambda^2}{6!}\Phi^6$. Replacing
this expression in (\ref{3}) we obtain \[ F(\alpha,\lambda)=
A(\alpha,\lambda)+B(\alpha,\lambda)+C(\alpha,\lambda)+D(\alpha,\lambda), \] where
\[
A(\alpha,\lambda)=\frac{1}{4}\int d^2x\, |x|^2\,
\langle\Phi^2(x)\Phi^2(0)\rangle_{\alpha},\]
\[
B(\alpha,\lambda)= -\frac{\lambda}{4!}\int d^2x\, |x|^2\,
\langle\Phi^2(x)\Phi^4(0)\rangle_{\alpha},\]
\[
C(\alpha,\lambda) = \frac{\lambda^2}{(4!)^2}\int d^2x\, |x|^2\,
\langle\Phi^4(0)\Phi^4(x)\rangle_{\alpha},\] and
\[
D(\alpha,\lambda) = \frac{\lambda^2}{6!}\int d^2x\, |x|^2\,
\langle\Phi^2(x)\Phi^6(0)\rangle_{\alpha}.\]

At this point we notice that $D(\alpha,\lambda)$, up to this order, contains only
tadpoles which, as explained above, were already considered in the renormalization of
$\alpha$. Then we must disregard this contribution in the present context. Now, in
order to illustrate the main features of the computation, we shall briefly describe
the evaluation of $A(\alpha,\lambda)$, which involves both analytical and numerical
procedures. In the above equations $\langle\ \rangle_{\alpha}$ means v.e.v. with
respect to the SG Lagrangian expanded up to second order in $\lambda$. From now on we
will decompose this Lagrangian into free and interaction pieces as \[ L_0 =
\frac{1}{2}(\partial_\mu\Phi)^2 + \frac{\alpha}{2}\,\Phi^2 ,\] \[ L_{int} =  -
\frac{\alpha\lambda}{4!}\,\Phi^4 + \frac{\alpha\lambda^2}{6!}\,\Phi^6 .\]

Using Wick's theorem and the well-known expression for the free bosonic propagator
$\langle\Phi(x)\Phi(0)\rangle_{0} = 1/(2\pi)K_{0}(\sqrt{\alpha}|x|)$ ($K_{0}$ is a
modified Bessel function of zeroth order), after a convenient rescaling of the form $x
\rightarrow \frac{x}{\sqrt{\alpha}}$, we obtain \begin{eqnarray}  \fl
A(\alpha,\lambda) = \frac{1}{2\alpha^2(2\pi)^2}\int d^2x\, |x|^2\, K_{0}^{2}(|x|) +
\frac{\lambda}{4\alpha^2(2\pi)^4}\int \int d^2x\, d^2x_{1}\, |x|^2\, K_{0}^{2}(|x_{1}
- x|)K_{0}^{2}(|x_{1}|) \nonumber
\\ \lo +\frac{\lambda^{2}}{\alpha^{2}(2\pi)^6}\int \int \int d^2x\, d^2x_{1}\, d^2x_{2}\,
|x|^2\, [\frac{1}{8}K_{0}^{2}(|x_{2} - x_{1}|)K_{0}^{2}(|x_{1}|)K_{0}^{2}(|x_{2} - x|)
\nonumber
\\ \lo + \frac{1}{6} K_{0}^{3}(|x_{2} -
x_{1}|)K_{0}(|x_{2}|) K_{0}(|x_{1} - x|)K_{0}(|x|) \nonumber
\\ \lo + \frac{1}{4}K_{0}^{2}(|x_{2} -
x_{1}|)K_{0}(|x_{1}|)K_{0}(|x_{2} - x|) K_{0}(|x_{2}|)K_{0}(|x_{1} - x|)].\nonumber
\end{eqnarray}

The first two terms are related to tabulated integrals \cite{table} yielding the first
order analytical expression $F(\alpha,\lambda)=(1+\lambda/(4\pi))/(12\pi\alpha^2)$
which can be shown to verify (\ref{3}) in a straightforward way. Concerning the
remaining second order terms we have three contributions corresponding to the
prefactors $1/8$, $1/6$ and $1/4$ in the last integrand. Due to its symmetry, the
first one can be analytically computed by repeatedly using \[ \int d^2x\, |x|^2\,
K_{0}^2(|x - y|) = \frac{2\pi}{3} +\pi |y|^2.\] The result is $2\pi^3$. The
computation of the other two contributions is more involved. In fact we could not find
analytical results in these cases. However we were able to considerably simplify these
multiple integrals in order to facilitate their numerical evaluation. Indeed, using
the Fourier transform of $K_{0}$ and the integral representation of the first kind
Bessel function $J_{0}$: \[J_{0}(|x|)= \frac{1}{2\pi}\int_{0}^{2\pi} d\theta\,  \exp
\left(-i|x|\cos\theta\right)
 ,\]
we obtain
\begin{eqnarray} \int \int \int
d^2x\, d^2x_{1}\, d^2x_{2}\, |x|^2\, K_{0}^3(|x_{2} -
x_{1}|)K_{0}(|x_{1}|)K_{0}(|x_{2} - x|)\nonumber \\ \times K_{0}(|x|) = 4(2\pi)^3\int
dr \int dp\, \frac{r\, p\, (1 - p^2) K_{0}^3(r)J_{0}(pr)}{(p^2 + 1)^5}.\nonumber
\end{eqnarray}
Using NIntegrate in the program Mathematica for the double integral in this expression
one obtains the value $0,04874$. Similar manipulations with the last contribution to
$A(\alpha,\lambda)$ led us to the following numerical results: \[  \int dr \int dp\
\int dk \frac {r\, p\, k\, (1 - p^2) K_{0}^3(r)J_{0}(pr)J_{0}(kr)}{(p^2 + 1)^4(k^2 +
1)^2} = 0,0169622 ,\] and \[  \int dr\, r^5\, K_{0}^2(r)K_{1}^2(r) = 0,08783 .\]

Putting all this together we have
\begin{eqnarray}   A(\alpha,\lambda) = \frac{1}{12\pi\alpha^2}(1+\frac{\lambda}{4\pi})+
\frac{\lambda^2}{\alpha^2(2\pi)^6}\, [\frac{1}{8}\,2\pi^3+\frac{1}{6}\,4\,(2\pi)^3\,
0.04874 \nonumber \\ + \frac{1}{4}\,(2\pi)^3\,(8\times 0.0169622-\frac{1}{8}\,
0.08783)] .\nonumber
\end{eqnarray} Working along the same lines with $B(\alpha,\lambda)$ and
$C(\alpha,\lambda)$ we find \[ \fl B(\alpha,\lambda)=
\frac{-32\lambda^2}{\alpha^2(2\pi)^34!}\int dr \int dp\, \frac{r\, p\, (1 - p^2)
K_{0}^3(r)J_{0}(p\, r)}{(p^2 + 1)^4}=\frac{-32\lambda^2}{\alpha^2(2\pi)^34!}\times
0,0501125 \] and \[ \fl C(\alpha,\lambda) = \frac{\lambda^2}{\alpha^2(2\pi)^34!} \int
dr\, r^3\, K_{0}^4(r)=\frac{\lambda^2}{\alpha^2(2\pi)^34!}\times 0,0754499 .\] All
these numerical values were confirmed by using Fortran.

Finally, inserting these results in the left hand side of (\ref{3}) we get \[
F(\alpha,\lambda)= \frac{1}{12\pi\alpha^2} + \frac{\lambda}{48\pi^2\alpha^2} +
\frac{\lambda^2}{4!(2\pi)^3\alpha^2}\left(\frac{3}{4} + 5.4 \times 10^{-6}\right).\]
Comparing this expression with the expansion in $\lambda$ of the right hand side of
(\ref{3}) one sees that they are equal up to first order and differ in a small
quantity (O($10^{-6}$)) up to second order in $\lambda$.

In summary, we have verified the validity of a quantitative prediction of CFT in the
context of the SG model. Taking into account the well-known bosonization identity
between the SG theory and the massive Thirring model (characterized by a coupling
$g^2$) \cite{Coleman} which takes place for $\beta^2/(4 \pi)= (1+g^2/\pi)^{-1}$, it
becomes apparent that our result implies that (\ref{1}) also holds in the strong
coupling limit of the Thirring model. This, in turn, allows us to improve the proof of
the validity of (\ref{1}) for the Baxter and Ashkin-Teller models which, up to now,
was restricted to the weak coupling limit \cite{Naon}.

\ack This work was partially supported by Universidad Nacional de La Plata (Argentina)
and CONICET (Argentina). M S thanks CICPBA (Argentina) for a Fellowship for Students.
C N is indebted to Bernard Jancovici for useful correspondence and stimulating e-mail
exchanges. The authors are also grateful to Marta Reboiro for helping them with
Fortran, and Vicky Fern\'andez and An\'{\i}bal Iucci for helpful comments.

\end{document}